# A new Look at the Electron Diffusion Region in Asymmetric Magnetic Reconnection

Michael Hesse[1,2], Cecilia Norgren[1], Paul Tenfjord[1], James L. Burch[2], Yi-Hsin Liu[3], Naoki Bessho[4], Li-Jen Chen[5], Shan Wang[4], Håkon Kolstø[1], Susanne F. Spinnangr[1], Robert E. Ergun[6], Therese Moretto[1], and Norah K. Kwagala[1]

[1]Space Plasma Physics Group, University of Bergen, Bergen, Norway
[2]Southwest Research Institute, San Antonio, Texas, USA
[3]Dartmouth College, Hanover, New Hampshire, USA
[4]University of Maryland, College Park, Maryland, USA
[5]NASA Goddard Space Flight Center, Greenbelt, Maryland, USA
[6]University of Colorado, Boulder, Colorado, USA

Corresponding author: Michael Hesse (michael.hesse@uib.no)

**Key Points:**

- Flow stagnation point and current density maximum are not necessarily collocated in asymmetric magnetic reconnection.

- The reconnection electric field sustains the electron current density and pressure also in asymmetric magnetic reconnection.

- Electron crescent distributions feature complex substructures related to different electron inflow regions.

**Abstract**
  A new look at the structure of the electron diffusion region in collisionless magnetic reconnection is presented. The research is based on a particle-in-cell simulation of asymmetric magnetic reconnection, which include a temperature gradient across the current layer in addition to density and magnetic field gradient. We find that none of X-point, flow stagnation point, and local current density peak coincide. Current and energy balance analyses around the flow stagnation point and current density peak show consistently that current dissipation is associated with the divergence of nongyrotropic electron pressure. Furthermore, the same pressure terms, when combined with shear-type gradients of the electron flow velocity, also serve to maintain local thermal energy against convective losses. These effects are similar to those found also in symmetric magnetic reconnection. In addition, we find here significant effects related to the convection of current, which we can relate to a generalized diamagnetic drift by the nongyrotropic pressure divergence. Therefore, only part of the pressure force serves to dissipate the current density. However, the prior conclusion that the role of the reconnection electric field is to maintain the current density, which was obtained for a symmetric system, applies here as well. Finally, we discuss related features of electron distribution function in the EDR.

**Plain-language summary:**
Magnetic reconnection is arguably the most important mechanism to release energy stored in magnetic fields explosively. Magnetic reconnection is believed to be the driver between as diverse a set of phenomena as solar eruptions, astrophysical radiation bursts, magnetic storms in near-Earth space, and the aurora. Quite amazingly, magnetic reconnection facilitates energy conversion over huge regions of space with size of many Earth radii by means of a tiny core region, the so-called diffusion region, with dimensions of a few to a few hundreds of kilometers. The delicate interaction between charged particles and electromagnetic fields in this central region enable the large-scale conversion of magnetic energy to particle energy to proceed. This paper presents a new look at the inner workings of this region for a fairly generic case of magnetic reconnection, which, among others, occurs at the interface between the Earth's magnetic field and the particle streams originating at the sun.

## 1. Introduction

Magnetic reconnection is arguably the most important energy conversion and transport process in collisionless plasmas. Magnetic reconnection is believed to be a key driver in astrophysical plasmas (e.g., Uzdensky, 2011), the mechanism behind solar eruptions (e.g., Antiochos et al., 1999), and it facilitates both energy entry into the magnetosphere as well as energy dissipation inside the magnetosphere (e.g., Hesse and Cassak, 2020). For these reasons, magnetic reconnection is also the ultimate engine behind many of the deleterious effects associated with space weather.

For both reasons of basic physical understanding as well as space weather-related applications reconnection has been a prime space physics research target. This began early on in solar research and continues today after the launch of the Magnetospheric Multiscale (MMS) mission, which provides unprecedented scientific evidence related to this fundamental process. Because of its orbital strategy, the MMS mission provided observations both of magnetopause reconnection, and of reconnection in the nightside, magnetotail current sheet. Of these two locations, the magnetopause is more challenging scientifically due to the intrinsic asymmetry of the reconnection inflow conditions, as well as the frequent presence of a guide magnetic field, i.e., the frequent occurrence of reconnection with inflow magnetic fields at angles other than 180 degrees with one another. Due to the importance of magnetopause reconnection for solar wind-provided energy entry into the magnetosphere – and perhaps also because of the intrinsic scientific challenge its complexity poses – magnetopause reconnection has been a prime focus for research both based on spacecraft observations and on theory and modeling (e.g., Russell and Elphic, 1979; Nykyrii and Otto, 2001; Cassak and Shay, 2007, 2009; Chen et al., 2016a; Shay et al., 2016; Genestreti et al., 2017; Liu et al., 2018).

MMS-based research has led to tremendous progress in understanding the physics of the inner core of the reconnection engine, the electron diffusion region (EDR). For example, MMS observations proved conclusively that predictions regarding crescent-like electron orbit signatures (Hesse et al., 2014, 2016) are correct (Burch et al., 2016; Burch and Phan, 2016). Hence, the laminar model for the EDR (Hesse et al., 1999), which applies directly in the tail (Nakamura et al., 2018), should in some form also apply at the magnetopause, even though there can be significant turbulence in the vicinity (e.g., Ergun et al., 2016). However, the tail results do not immediately transfer to the magnetopause, and more work needs to be done to understand the detailed structure of the EDR, as well as how electron current dissipation happens here. This paper reports on progress in researching these questions. Specifically, we will present modeling results pertaining to current dissipation and electron heating processes around two key locations in the EDR: the flow stagnation point, and the current density maximum, which, surprisingly, do not coincide in our simulation. After comparing our new results to prior results for symmetric magnetic reconnection, we will proceed to an in-depth study of electron distribution function to shed light on the kinetic foundations of the current dissipation and heating results.

The paper is organized in the following way: section 2 presents the model as well as the simulation setup, and it provides an overview of the system evolution. Section 3 provides a closer look at the structure of the EDR as well as the composition of the electric field. Section 4 presents results of current continuity and energy conservation for stagnation point and current maximum, and section 5 augments these results with an analysis of electron distributions functions in this area. Finally, section 5 presents a summary as well as conclusions.

## 2. The model and evolution overview

For ease of the analysis, we introduce normalized quantities. We simulate a proton-electron plasma with a mass ratio $m_i/m_e=100$. The particle mass is normalized to the proton mass, the magnetic field to a typical value in the inflow region ($B_0$), and the density to a typical density ($n_0$) in the current layer. Building upon these units, time is normalized to the inverse of the ion cyclotron frequency $\Omega_i = eB_0/m_i$, lengths are measured in units of the ion inertial length $d_i=c/\omega_{pi}$, and the velocity unit is the proton Alfven speed $v_A = B_0/\sqrt{\mu_0 m_i n_0}$. The ratio between the electron characteristic frequencies is set to $\omega_{pe}/\Omega_e=2$. The simulation employs our proven simulation code (e.g., Hesse et al., 2018), here in a 2.5-dimensional configuration. For the purpose of the present investigation, we employ 3200x3200 cells, and a total of $7.2 \times 10^{10}$ particles. Ion and electrons are initialized as Maxwellians with density, temperatures and drift velocities corresponding to the initial conditions below. Our choice of coordinate system has $x$ as the initial magnetic field direction (corresponding to the L direction often employed in space observations), $y$ as the direction of initial current flow (corresponding to M), and $z$ as the reconnection inflow direction (corresponding to N).

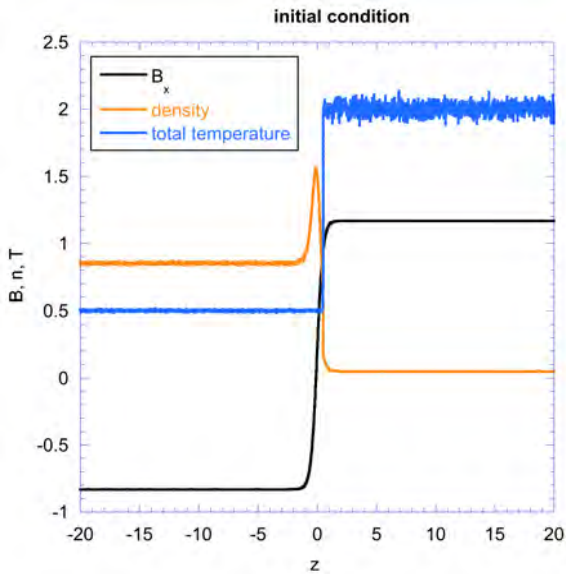

Figure 1 – Profile of the initial condition. The magnetospheric side, with higher temperature, larger magnetic field strength, and lower density is located at $z>0$. The field reversal is initially filled by magnetosheath-like plasma.

The initial condition models an asymmetric configuration with a temperature gradient. The magnetic field is initialized to:

$$B_x = \tanh(z/l) + a_1 \qquad (1)$$

with a current sheet thickness $l=0.5$, and an asymmetry of $a_1=1/6$. The out-of-plane magnetic field is set to zero in the initial condition, and an X-type perturbation is added to the magnetic field to speed-up the formation of a reconnection region. The initial total pressure is:

$$p = 0.1 + 0.5(1+a_1)^2 - 0.5 B_x^2 \qquad (2)$$

The density initialization takes into account the desired temperature variation in the following way:

$$n = 0.5p \text{ for } z > l \tag{3a}$$

$$n = 2p \text{ for } z < l \tag{3b}$$

This choice implies a temperature variation for a combined value of $T=2$ on the upper side (corresponding to the magnetosphere) to $T=0.5$ on the sheath side. In order to reduce fluctuations generated by the lack of a kinetic equilibrium, the initial current sheet population is initialized like the magnetosheath. This initial condition is displayed in Figure 1.

The particle table is split into two groups: one representing the magnetospheric population, and the other the sheath and the initial current layer. This choice permits a ready assessment of particle origin. The ratio of ion-to-electron temperature is chosen to be $T_i/T_e=5$, and the physical system size is $L_x \times L_z = 102.4 d_i \times 51.2 d_i$. The system is integrated using a time step of $\omega_{pe}dt=0.5$ until just before the analysis time, after which the time step is changed to $\omega_{pe}dt=0.01$ to guarantee extremely accurate integration of particle trajectories in the simulation.

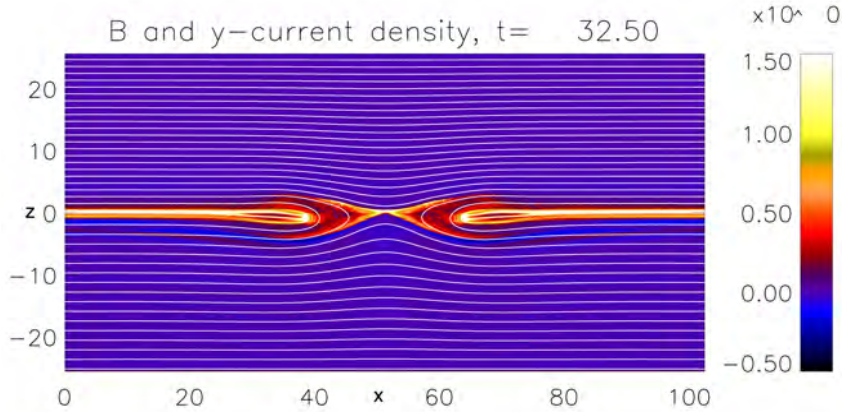

Figure 2 – State of the simulated system at the time of investigation. The plot shows in-plane magnetic field (white lines)

Figure 2 shows the simulation at $t=32.5$, the time of analysis. We find the patterns typically associated with asymmetric magnetic reconnection: stronger current density on the magnetospheric, i.e., upper, side, and a more pronounced bulge of the reconnected current sheet on the magnetosheath (lower) side. There is evidence of current filamentation on the magnetospheric side separatrix, a likely indication of electron holes and electrostatic turbulence similar to what is found in symmetric models (e.g., Divin et al., 2012). In the following, we will take a closer look at the structure of the electron diffusion region.

3. **Structure of the electron diffusion region**

A close-up of the electron diffusion region, showing magnetic field, out-of-plane current density, and in-plane electron flow, is shown in Figure 3a. A number of features are apparent: we find the expected electron flow through the X-point toward the magnetospheric side, with a flow stagnation point clearly separated from the X-point. The flow stagnation point is not located at the current density maximum, but rather significantly displaced further into the magnetospheric region. In the regions around the separatrices we find strong electron flows, which are largely aligned with the in-plane magnetic field. These flows are primarily providing the current density associated with the reconnection-generated out-of-plane magnetic field, which is shown in Figure 3b, and not with the transport of in-plane magnetic flux. In addition to this bipolar magnetic field, the vectors in F3b denote the component of the in-plane electron flow, which is perpendicular to the in-plane magnetic field, and thus related to the transport of magnetic flux if the electrons are frozen into the magnetic field. We find indeed that the flux transport velocity field overall has considerably smaller amplitudes than the total electron flow velocity.

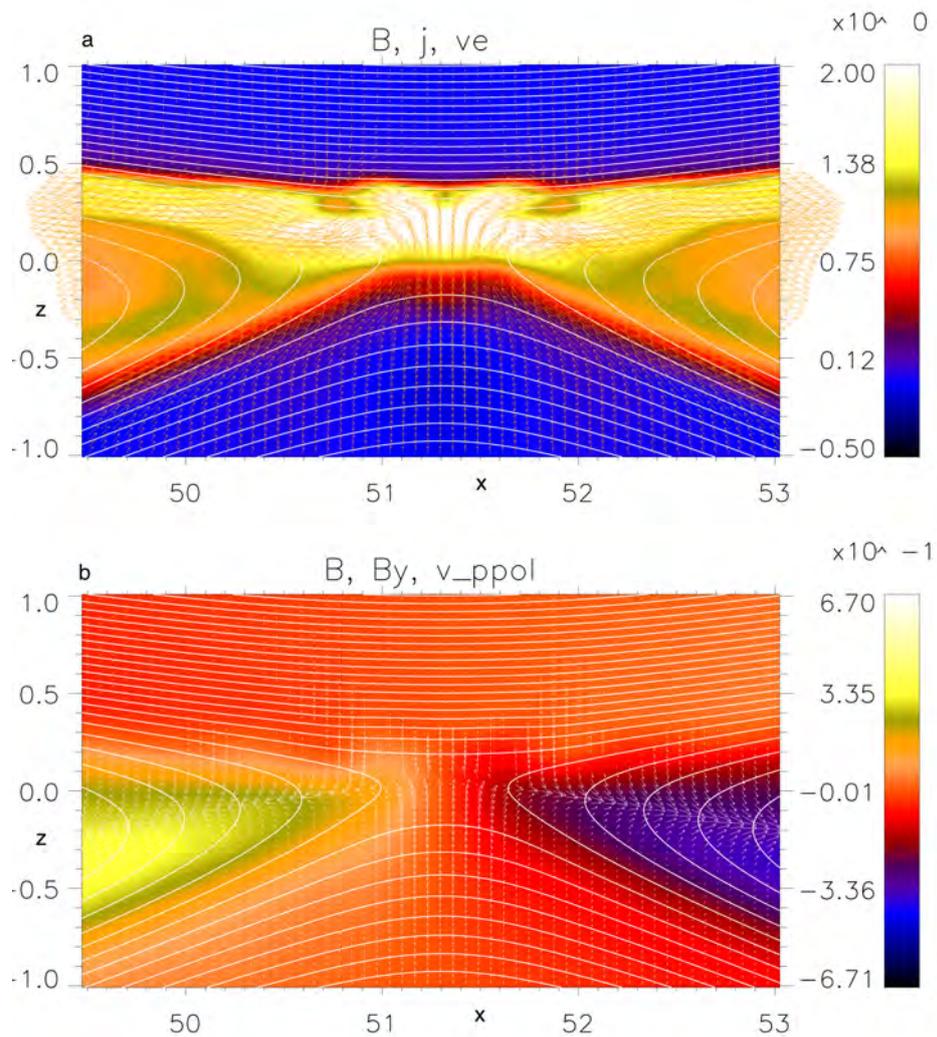

Figure 3 – (a): blowup of the in-plane magnetic field, current density, and in-plane electron flow vectors. The panel shows a clear separation of the in-plane stagnation point from the maximum of the current density. (b): in-plane magnetic field lines, out of plane magnetic field $B_y$ (color), and the perpendicular components of the in-plane electron flow, signifying magnetic flux transport.

We now take a closer look at the electron diffusion region, by means of a cut in the $z$-direction at the location of the X-point: $x=51.325$. The left panel of Figure 4 shows a set of key physical quantities along the $z$-direction. We find the X-point ($B_x=0$) to be located at $z=0.01$, and electron flux in the $y$-direction to be peaking on the magnetospheric side, approximately at $z=0.19$. The y-component of the electric field is nearly constant in this region, indicative of a reasonably good steady state. The z-component of the electron flow velocity penetrates deep into the magnetospheric side, featuring a stagnation point ($v_{ex}=v_{ez}=0$) at $z=0.36$, well separated from the electron current density peak. The physical separation of the stagnation point and the current peak adds additional complexity to the asymmetric electron diffusion region. The Hall-type electric field $E_z$ is found to be very strong, more than a factor of ten larger than the reconnection electric field. Finally, we see that the nongyrotropic electron pressure tensor component $P_{yze}$ features a strong gradient at the current peak and should hence be expected to provide a major contribution to the reconnection electric field. At the stagnation point, however, no such gradient is apparent. Hence pressure electric field contributions here must result from x-derivatives of $P_{xye}$.

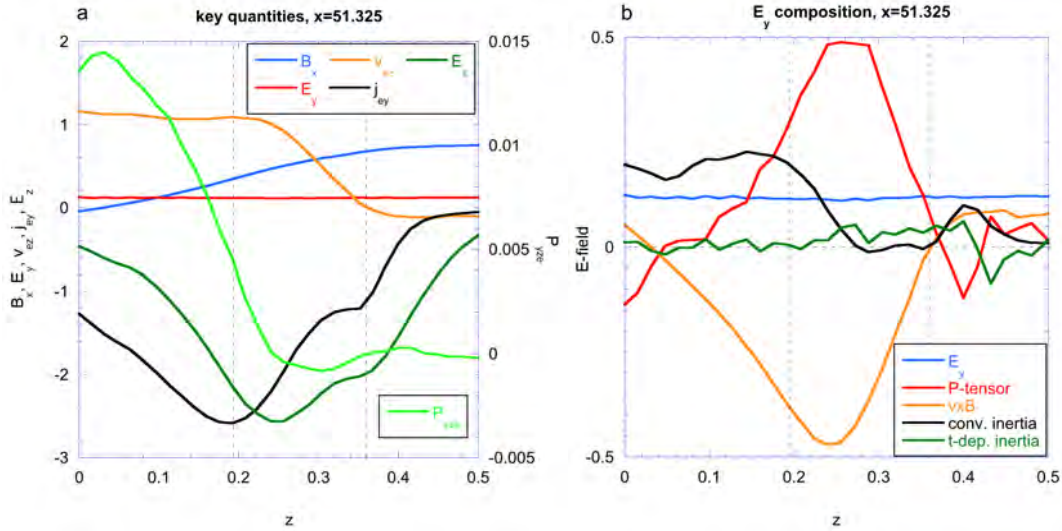

Figure 4 – (a): plot of the reconnecting magnetic field $B_x$, the electron flow velocity $v_{ez}$, the normal electric field $E_z$, the reconnection electric field $E_y$, the y-directed electron flux $j_{ey}$, and the nongyrotropic pressure component $P_{yze}$, along a line connecting X-point and flow stagnation point, (b): decomposition of the reconnection electric field along the same line. The dashed lines indicate the approximate location of the current density maximum (left) and the flow stagnation point (right).

Figure 4b presents an analysis of the y-component of Ohm's law as expressed by the electron momentum equation:

$$E_y = -\frac{1}{en_e}\left(\frac{\partial P_{yze}}{\partial z} + \frac{\partial P_{xye}}{\partial x}\right) - v_{ez}B_x + v_{ex}B_z - \frac{m_e}{e}\vec{v}_e \cdot \nabla v_{ey} - \frac{m_e}{e}\frac{\partial}{\partial t}v_{ey} \qquad (4)$$

Each term of (4) is plotted in order in F4b. We find a rather complex role of the individual terms. The pressure tensor term (first term on the RHS of (4)) actually subtracts from the reconnection electric field at the X-point, but provides, by far, the largest positive contribution around the current density maximum. It provides the majority of the electric field at the stagnation point; the rest is provided by the last term in (4). The convection (vxB) term provides a large negative contribution primarily around the current maximum, which results from the electron motion against the ExB drift direction. We will return to this feature later. Not shown here is that the electrons become fully frozen into the magnetic field only at approximately $z>1$ and $z<-1$. Similar to earlier investigations (Hesse et al., 2014), the convective inertia term provides the bulk of the reconnection electric field at the X-point.

Finally, there are some smaller time-dependent effects, which complete the decomposition of the reconnection electric field in this region. We will ignore these in the following discussion.

## 4. Current and energy balance

In symmetric systems, the electron diffusion region is the site of an intricate balance between energization and dissipation of current and thermal energy (Hesse et al., 2018). At least in principle, a similar balance should also exist in asymmetric reconnection, but this has, to-date, not been demonstrated. The purpose of this section is to present the results of a corresponding investigation for asymmetric reconnection.

In the electron diffusion region, the ion current is negligibly small compared to the current carried by the electrons. Similar to Hesse et al. (2018), the current balance can therefore be expressed in form of an appropriate modification of the electron momentum equation:

$$-\frac{\partial}{\partial t}en_e v_{ey} = \frac{e^2 n_e}{m_e}E_y + \frac{e^2 n_e}{m_e}v_{ez}B_x - \frac{e^2 n_e}{m_e}v_{ex}B_z + \frac{e}{m_e}\left(\frac{\partial P_{yze}}{\partial z} + \frac{\partial P_{xye}}{\partial x}\right) + e\nabla \cdot n_e \vec{v}_e v_{ey} \quad (5)$$

The left side of this equation is the time derivative of the electron current density, the first term of the right-hand-side is the electric field force term, followed by the Lorentz force contribution (terms two and three), the pressure term, and the convection term. These terms balance each other along the z-cut presented above; however, it is less clear which terms dominate if a larger region is being considered, and whether any dominance is generic.

Another consideration of critical importance to the structural maintenance of the electron diffusion region is the internal energy balance. A loss of internal energy, or, equivalently, a pressure reduction, would lead to a collapse of the EDR and hence the current layer, in the same way as an oversupply of thermal energy would lead to an expansion. The time evolution of the trace of the electron pressure tensor $p = Tr(\vec{\vec{P}})/3$ is, which is proportional to the electron thermal energy, is governed by:

$$\frac{\partial p}{\partial t} = -\nabla \cdot (\vec{v}p) - \frac{2}{3}\sum_l P_{ll}\frac{\partial}{\partial x_l}v_l - \frac{1}{3}\sum_{l,i}\frac{\partial}{\partial x_l}Q_{lii} - \frac{2}{3}\sum_{\substack{l,i \\ l \neq i}} P_{li}\frac{\partial}{\partial x_i}v_l \quad (6)$$

On the right-hand side of this equation, the first two terms are convection-compression terms, which become MHD-like if the second term only involves isotropic pressure tensor components. The divergence of the trace of the heat flux tensor (third term) can be interpreted as a correction to the first two transport terms to account for more complex particle distribution functions (Hesse et al., 2018). We will therefore treat the first three terms collectively and refer to them as generalized transport terms. Finally, the last term involves shear-type derivatives of the electron flow velocity, and off-diagonal terms of the electron pressure tensor, which can be due to nongyrotropy. We will therefore refer to the contributions from this term as "quasi-viscous."

In order to investigate generic balances in both current and internal energy, we integrate the individual terms of (5) and (6) over a family of rectangles, defined as

$$x_0 - 5d \leq x \leq x_0 + 5d \quad (7a)$$

$$z_0 - d \leq z \leq z_0 + d \quad (7b)$$

The half-thickness d is varied between *d=0.03* and *d=0.1*, the latter corresponding approximately to an electron inertial length. The integration region is centered about a point of interest, located at *x=$x_0$* and *z=$z_0$*. As we have seen in the preceding section, the relatively simple geometry of symmetric systems, where X-point, current peak, and stagnation point usually coincide, is changed considerably in asymmetric reconnection. Therefore, it is appropriate to investigate these balances separately around the stagnation point, and around the maximum of the current density. The results are presented separately in the following two subsections.

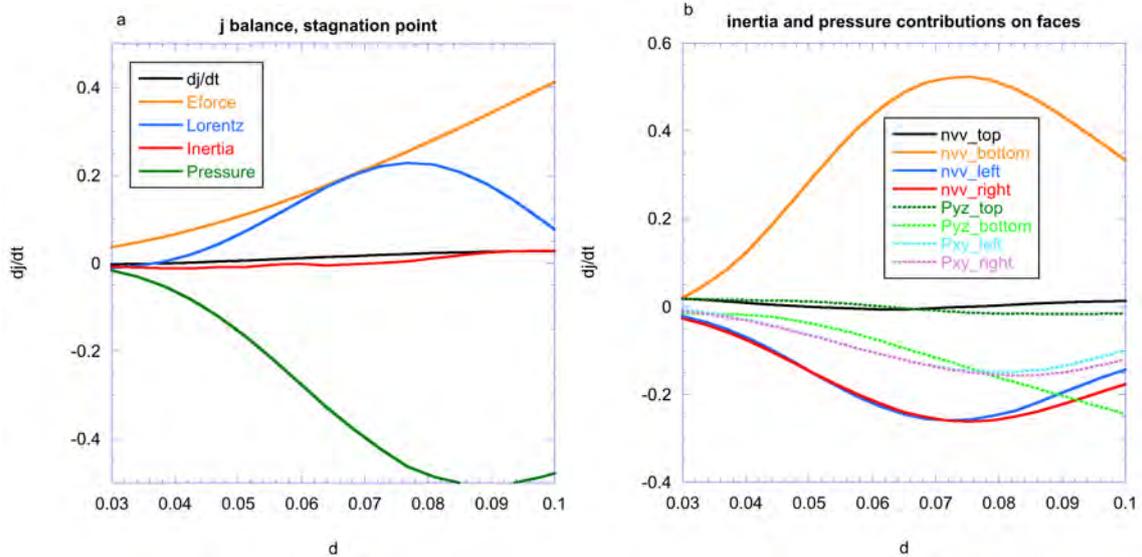

Figure 5 – Results of the current balance integration. (a): current balance around the stagnation point. The panel demonstrates that the pressure force works consistently to reduce the integrated current, whereas the electric field balances this reduction. The Lorentz force effect can be explained by a generalized electron diamagnetic drift. (b): Effects of inertial and pressure at each side of the integration rectangle (denoted by top, bottom, left, and right). There is substantial current convection from below, which is completely balanced by current convection out of the volume at the sides. Of the pressure terms, the panel shows that $P_{xye}$ at each side and $P_{yze}$ at the lower boundary are equally important.

**4.1 Stagnation point**

As the key critical point between reconnection in- and outflows, the stagnation point plays a special role in any reconnection configuration, and hence a natural location to investigate the electron current and energy balance. Results for the current balance integration about the stagnation point location *$x_0$=55.33, $z_0$=0.36*, are displayed in Figure 5. The left panel shows that throughout the range of sizes of the integration region, the electric field acts as a current generator. An apparent second contributor to current generation is provided by the Lorentz force term, while the inertia term does not contribute in a significant way to the overall current balance. The integrated current density does not change significantly in time because the pressure dissipation is almost perfectly balanced by current generation. We point out that an alternative way to look at this balance is to interpret the Lorentz force term as the result of a generalized diamagnetic drift generated by the divergence of the nongyrotropic pressure terms. In this view, only the sum of pressure and Lorentz force terms describes the dissipative effect balancing the electric field acceleration.

Figure 5b shows contributions at the integration area sides from the divergence terms in (5). We see that the relatively quiet magnetospheric inflow region does not impact the current balance by either thermal or convective effects. This is very different for the lower face and the two side faces. Here, and matching the flow patterns in F3, we find strong current transport into the integration box from below, and out to both sides. The transport into the volume is nearly exactly balanced by transport out, leading to the negligible net contribution in F5a. As for the thermal, i.e., pressure contributions, we find, for most of the integration area sizes, equal contributions at each of the left and right faces, and at the lower face. This result indicates that, around the stagnation point, the $x$-derivative of $P_{xye}$ can be more important than the $z$-derivative of $P_{yze}$.

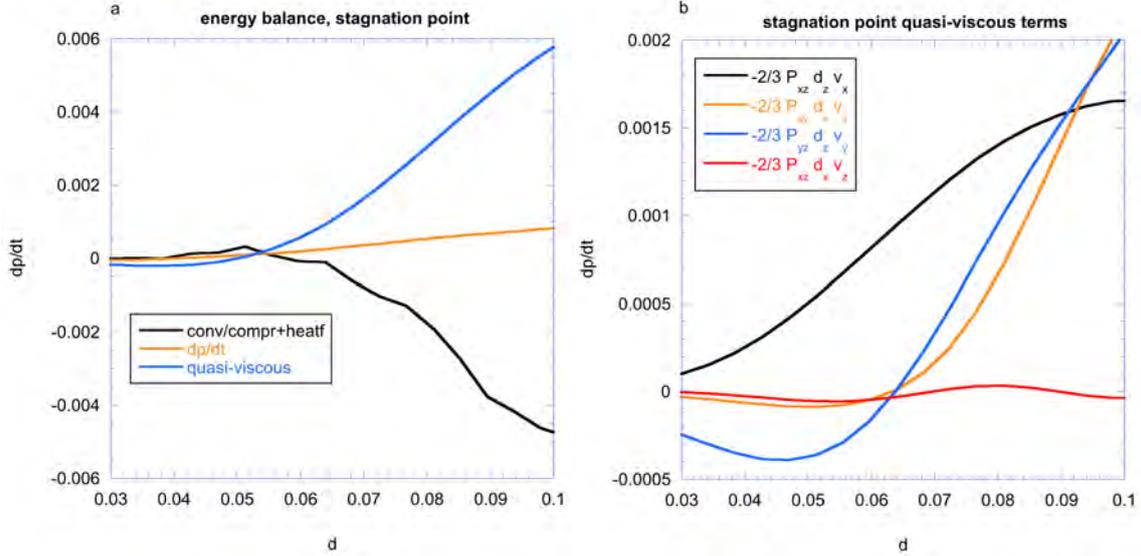

Figure 6 – Results of the energy balance integration. (a): energy balance around the stagnation point. We see the effect of quasi-viscous heating balancing the energy reduction by the combination of convective effects. (b): analysis of the role of the individual quasi-viscous contributions. For small integration sizes, there is little net effect, after which there is a clear heating role of all gradients associate with the main electron flows. The two pressure terms, $P_{xye}$ and $P_{yze}$, which play a key role in current reduction, also feature prominently as heating contributors.

The energy balance around the stagnation point is displayed in Figure 6. The left panel shows that there is very little net contribution from either the combination of compression-convection and heat flux terms, or from the quasi-viscous terms, if the integration region is small. After approximately $d=0.05$ we see the expected pattern emerge: heating by quasi-viscous terms balanced by a net energy reduction due to energy convection. Given that $P_{xye}$ and $P_{yze}$ play key roles in dissipating the current density and hence the kinetic energy associated with it, it would be a reasonable expectation that these terms also contribute to electron heating. Figure 6b shows that this expectation is correct, at least for the larger sizes of the integration rectangles. For these larger sizes both terms contribute about equally, whereas for smaller sizes the term involving $P_{yze}$ actually leads to a negative contribution, due to the positive sign of $P_{yze}$ on a positive $z$-gradient of the electron flow velocity in the $y$ direction. A somewhat surprising result is the contribution due to the product of $P_{xze}$ and $\partial/\partial z\, v_{ex}$. The importance of this term indicates that mixing on the rather sharp gradient of the electron flow outflow can also contribute to overall heating.

In summary, we find around the stagnation point a more complex situation than in symmetric reconnection. Common to both is the dominance of thermal, i.e., nongyrotropic pressure-based, current dissipation, and quasi-viscous heating. However, we here find substantial current convection toward and away from the stagnation point, and nongyrotropic pressure enabling electron flow penetration into the magnetospheric-side magnetic field against the *ExB* drift. Heating due to nongyrotropic pressure is also more complex than in symmetric systems, where heating appears predominantly caused by the *y-z* component of the pressure tensor. This heating is also found here, but only for larger integration areas, whereas heating is also found to be based on pressure nongyrotropies at sharp gradients of the (mostly field-aligned) electron outflow.

**4.2 Current density peak**

Based on our expectation that reconnection should, in the EDR, first and foremost be a current dissipation process, we expect that the region around the current density peak between the X-point and the flow stagnation point should be particularly interesting. We hence conduct the same kind of analysis as above, but this time the family of integration regions is centered on *$x_0$=55.33, $z_0$=0.20*, the approximate location of the current density peak. Figure 7 displays the results, in the same format as F5. The current balance shows similarities: the integrated current density features very little time dependence, the electric field acts to increase the current density together with the Lorentz-force term. This increase is countered by thermal dissipation, and, different from above, a net convective removal of integrated current density. We refer to this current convection effect as nondissipative, because convection in itself does not involve any thermalization. It is larger than the sum of the Lorentz and pressure terms. This sum could be interpreted as the effective current dissipation effects. We therefore find that asymmetric magnetic reconnection can feature both dissipative (i.e., thermal) and nondissipative contributions to the current balance with the acceleration force of the reconnection electric field.

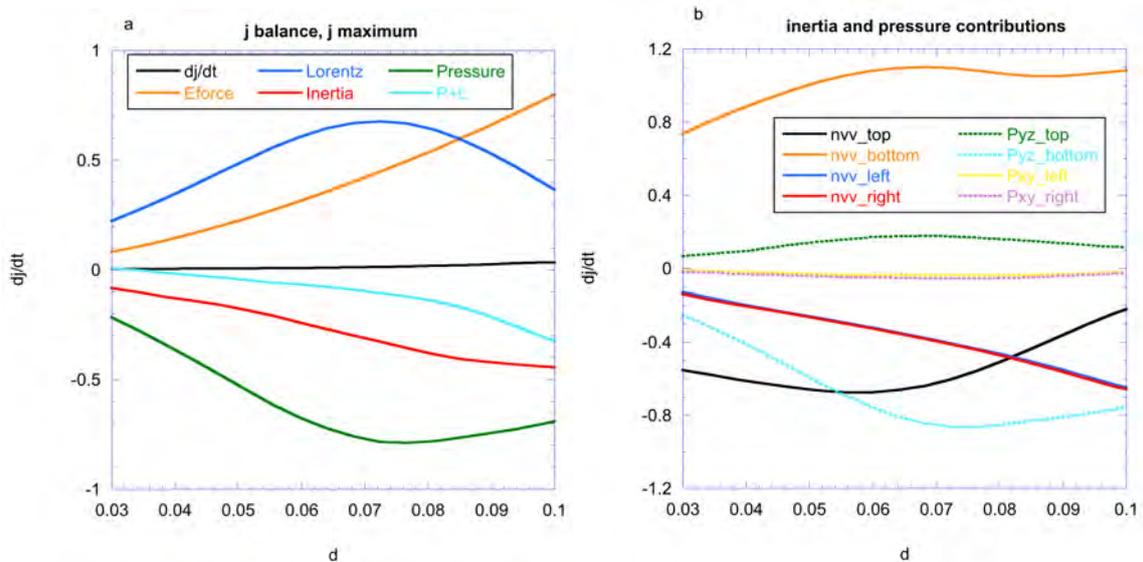

Figure 7 –(a): current balance around the current density peak. The panel demonstrates that the current density is reduced by the net effect of pressure and Lorentz-force (P+L), a well as through convection out of the volume (inertia). The latter effect is nondissipative and quite different from the scenario around the stagnation point. (b): Effects of inertial and pressure at each side of the integration rectangle. There is substantial current convection from below, which is over-compensated for by current convection out of the volume at the sides and at the top. Of the pressure terms, the panel shows that, contrary to the stagnation region, $P_{yze}$ at the lower boundary is the dominant contribution.

Figure 7b breaks out the pressure tensor and convection distributions at the individual sides of the integration region. The pressure tensor contributions show a rather dramatic dominance of $P_{yze}$ at the bottom boundary, with a small positive contribution at the top. Here and different from the area around the stagnation point, the $P_{xye}$ contributions at the left and right boundaries are negligibly small, indicating that these terms are unimportant for dissipation around the current density maximum. The convective current balance contribution features a major current supply from the lower boundary, and strong transport away from the integration region at the top and side boundaries, largely consistent with the in-plane flow patterns visible in F3. There is a net loss due to the convective transport, which establishes an additional requirement for electric field acceleration to maintain the current density.

The balance between quasi-viscous heating and energy transport around the current maximum (integration of (6)) is shown in Figure 8. Here we find, for rectangle sizes $d$ less than about 0.075, the expected result: clear heating by quasi-viscous terms, which balances the combination of compression, convection, and heat flux. Figure 8b shows that this heating is dominated by the product of the $P_{yze}$ component of the pressure and the $z$-derivative of the $y$-component of the electron flow velocity. This result is consistent with the dominance of this pressure term in the current balance, as well as the dominance of the $y$-directed electron flow over the in-plane electron flow components in this region. Gradients of the in-plane flow velocity are small for small-to-moderate integration regions. Therefore, we do not find heating associated with in-plane flows like we found around the stagnation point. However, we do see that this balance can change for the largest integration regions, where quasi-viscous terms can also add negative contributions in the energy balance. This feature is quite unique to collisionless plasmas; in collisional plasmas viscous processes are usually tied to heating.

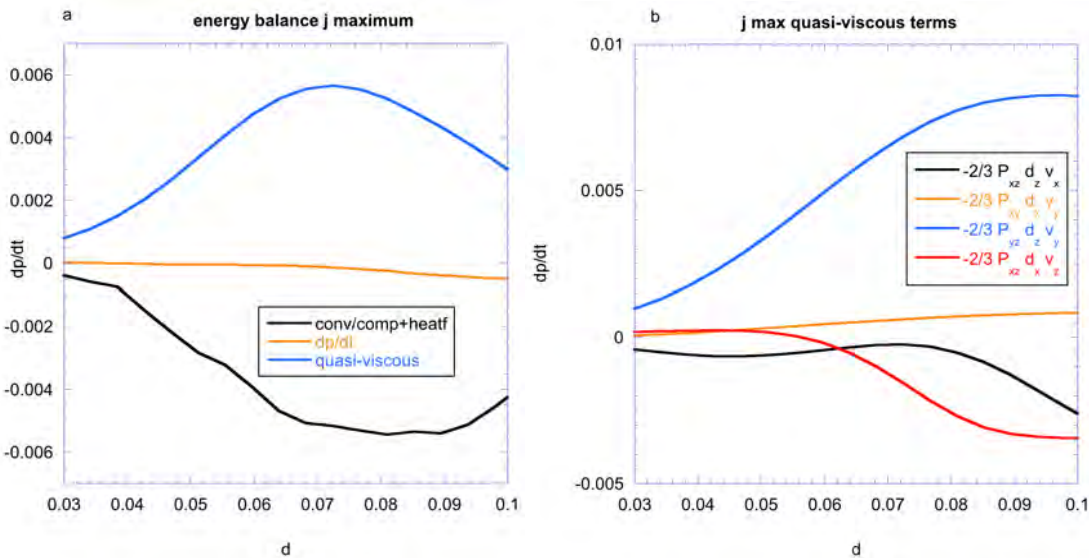

Figure 8 – (a): energy balance around the current density peak. We see also here quasi-viscous heating balancing the energy reduction by the combination of convective effects. (b): analysis of the role of the individual quasi-viscous contributions around the current density peak. The term involving $P_{yze}$, which plays the dominant role in current reduction, also dominates electron heating.

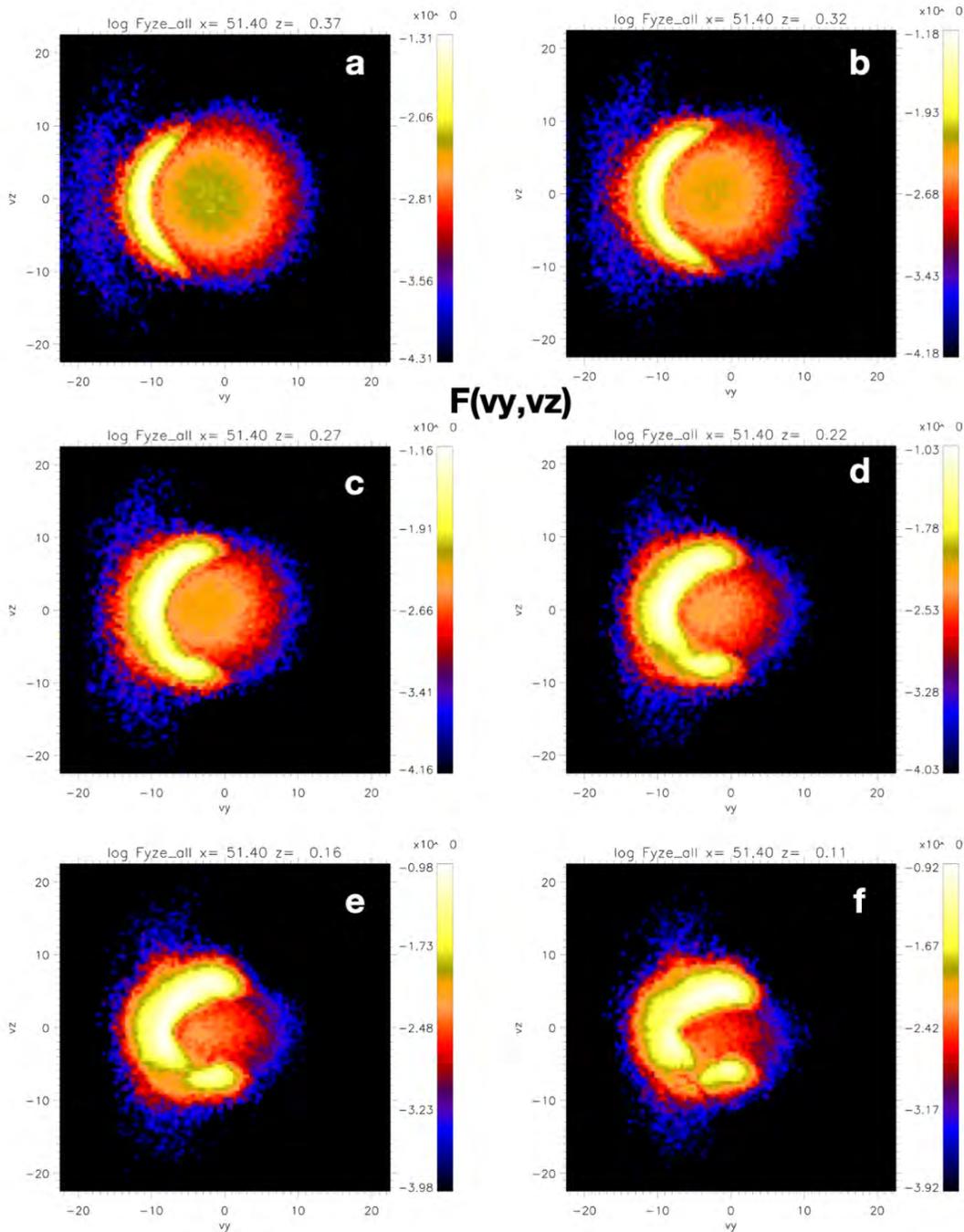

Figure 9 – logarithmic plots of the reduced distribution functions $F(v_y, v_z) = \int dv_x\, f(v_x, v_y, v_z)$ at different $z$-positions (indicated in the panels) along the line connection stagnation point and X-point.

Hence, we find that the current density peak has many features similar to what we know from symmetric systems. Nongyrotropic pressure effects play a key role in dissipating the local current, and the $P_{yze}$ component plays a substantial role in the dissipation as well as the electron heating. In addition, a key feature is rather different from symmetric systems: a dominant convective current transport, which, at least in principle, is not dissipative. However, the current density loss associated with it still needs to be balanced by the reconnection electric field. In the following section, we will investigate a select set of electron distribution functions in order to understand how the features identified here relate to electron structures in phase space.

## 5. Distribution function perspective

Our research of electron distributions is divided into two parts: reviewing the variation of the reduced distribution function $F(v_y,v_z)$ as a function of $z$ on a line, which connects the stagnation point and current maximum, and reviewing the variation of $F(v_x,v_y)$ as a function of $x$ at $z$-locations close to those of the stagnation point and the current maximum. The former is related to the aforementioned current transport in the $z$ direction, whereas the latter are related to current transport in the $x$ direction. We begin with the former

Figure 9 shows the reduced distribution $F(v_y,v_z)$ at six different locations in z along the line connecting the stagnation point and current maximum. We see a hotter and more gyrotropic distribution of magnetospheric origin, superposed on crescent-type distributions with considerably higher phase space density. The distributions are similar to what has been shown before for similar configurations (Bessho et al., 2016, 2017; Chen et al., 2016b), but there are also some notable differences. One particularly interesting one is the existence of what appears to be a second crescent in F9a, i.e., at the stagnation point. We will see below that this crescent is actually a triple structure. Particle tracing shows that this outer structure is composed of accelerated particles originating in the magnetospheric part of the models, whereas the bright feature of the main crescent is, as usual, due to bouncing particles of magnetosheath origin, which are accelerated in a very strong $E_z$ layer on the magnetospheric side. Moving down from the stagnation point, we see that the crescents develop a very strong asymmetry similar to what was found earlier (e.g., Chen et al., 2016a). We note here that this asymmetry leads to an average particle motion in the positive $z$ direction. Unlike in F9a, we see that the combination of partial crescents and magnetospheric population develop increasingly larger asymmetries in $v_z$, which lead to the larger values of $P_{yze}$ seen in F4a.

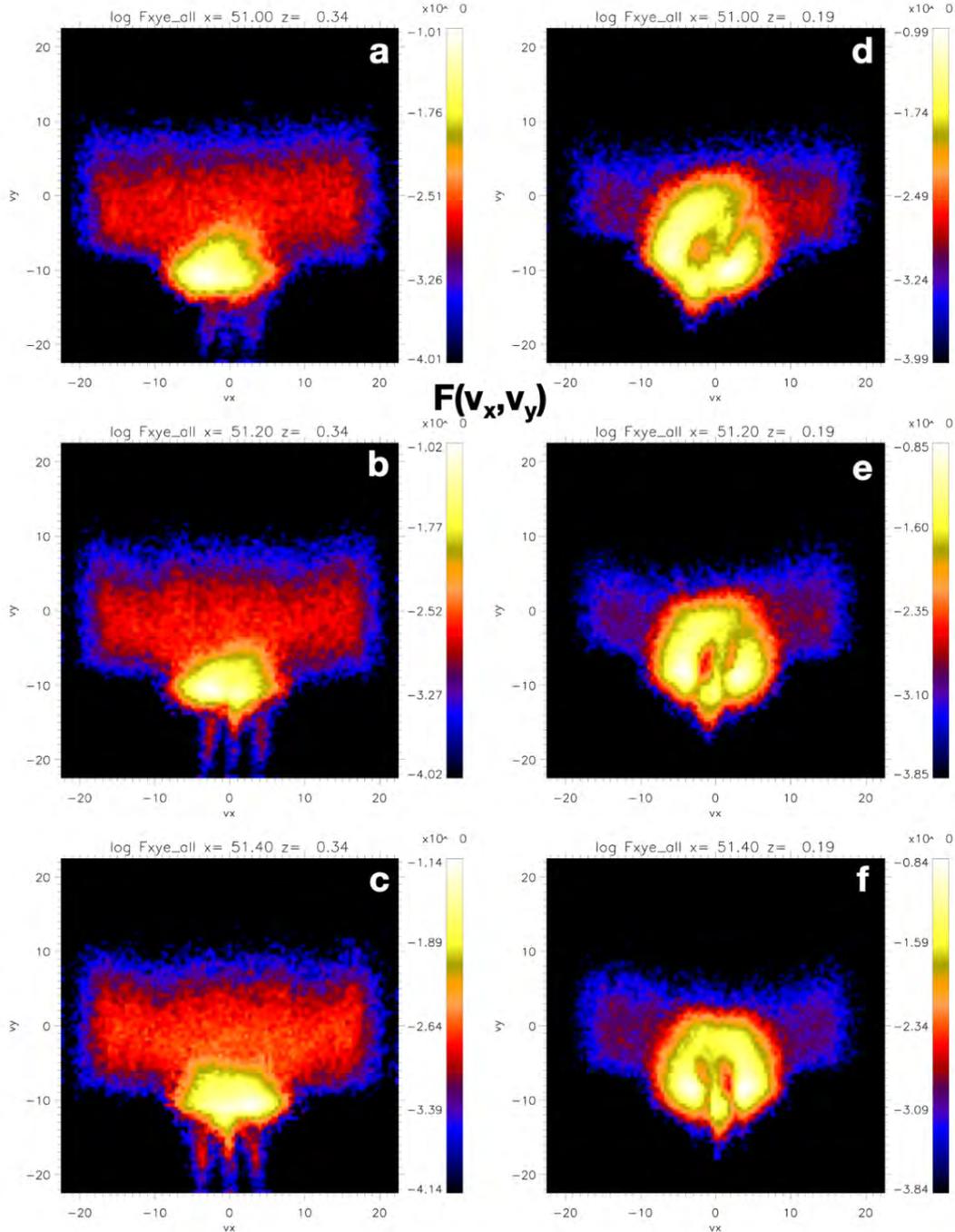

Figure 10 – logarithmic plots of the reduced distribution functions $F(v_x, v_y) = \int dv_z\, f(v_x, v_y, v_z)$ at different $x$ and $z=0.34$ (panels (a)-(c)), as well as at $z=0.19$ (panels (d)-(f)). Panels (a)-(c) display the cylindrical distribution of the magnetospheric inflow distribution, together with high space density crescents, and crab leg-like higher energy features. At $z=0.19$, the main crescents have devolved into complex, multi-pronged, phase-space structures.

The variation of $F(v_x,v_y)$ with $x$ for two different $z$ locations is shown in Figure 10. Approaching the stagnation point from the left, panels a-c show distributions composed of a hot, reasonably gyrotropic distribution with significantly larger parallel than perpendicular temperature, combined with a bright crescent feature, which is increasingly displaced in the negative $v_x$ direction for increasing distance from the stagnation point. In addition, we notice a crab leg-like feature at higher energies, which leads to the higher energy crescent described above when seen from the side. These crab legs are actually the line-of-sight integral of three individual crescents generated by particles from the magnetospheric side, which execute multiple bounces in the field reversal and normal electric field. The bulk motion, however, is clearly controlled by the much larger phase space density of the main crescents. We see furthermore that the increasing shift of the crescents in $v_x$ generates an increasing asymmetry of the distribution, which, in the center of mass frame, leads to significant values of $P_{xye}$.

This, relatively simple, geometry is replaced by substantial complexity at the $z$-height of the current density peak. We find that the crescent has devolved into a complex, three-dimensional structure, which, at the current density maximum features three different high-energy fingers (F10e). The left and right of these fingers are formed by particles of magnetosheath origin with finite initial velocity in $x$, which originates in the inflows just outside of the magnetosheath-side separatrix. This inflow speed is approximately preserved, whereas the particle gets energized by $E_z$ and, to a smaller degree, also by the reconnection electric field itself. Therefore, the outer two fingers are images of the electron inflow on the magnetosheath side. As particle tracing shows, the middle finger is composed of magnetosheath particles, which execute one additional bounce across the field reversal, and therefore gain additional energy by a longer drift in the reconnection electric field.

Moving to the left of the current maximum leads to a rotation and increasing smearing of the three-finger structure of the current maximum. While it appears evident that the phase space density weight shifts increasingly toward negative $v_x$, it is less clear what the effect of this distortion is on the pressure tensor component $P_{xye}$, which, as we know from the analysis above, does not appear to play a significant role in the current balance.

We conclude the distribution function analysis by a look at the three-dimensional structure of the electron distributions at the stagnation point and at the location of the current density maximum. These two distributions are shown in Figure 11. The stagnation point distribution, shown in the left panel, features the relatively broad main crescent, which does not exhibit any discernable substructure in the $v_x$ direction. Also visible are the three crab leg features, the right one of which (positive $v_x$) is also displayed in the vertical plane. This vertical plane plot proves that the crab legs are actually three distinct, higher-energy crescents. Finally, the inflowing, magnetospheric population is relatively featureless, spring roll-like about the magnetic field, as already described above.

The three-dimensional view of the distribution at the current density peak reveals even more complexity than what is evident from the reduced distributions. We see that the magnetospheric populations begins to develop structure along the $v_x$ directions. More dramatically, we notice that the outer two (in $v_x$) bright finger-features, which are caused by electrons with finite $v_x$ as noted above, are, in fact, individual, crescent-like phase space shapes. There is some indication of electrons with even higher energies, but the number of particles here is too small to make a conclusive assessment of their origin. Even though not shown here, the bright center structure is the intersection in the $v_z=0$ plane of a third crescent-like shape.

In summary, our investigation extends previous studies of similar distributions (Chen et al., 2016b, Bessho et al., 2017, 2017) by several new features. These features should, in principle, be observable by spacecraft measurements. Whether they are in fact observable depends both on instrument sensitivity and on angular resolution in velocity space.

## 6. Summary and conclusions

Reconnection in asymmetric systems remains a very challenging object of research. The comparatively simple symmetric reconnection geometry, where X-point, current density peak or saddle point, and flow stagnation point are approximately coincident, gets distorted considerably in asymmetric systems. In particular, we know that X-point and the flow stagnation point are spatially separated (Cassak and Shay, 2007, 2009), and the current density peak is also to be found on the high-field side of the X-point. The relation between current density peak and stagnation point remains unclear, but it has been speculated that they may coincide (Hesse et al., 2014).

We here reported on new research into the asymmetric electron diffusion region in a system without an initial guide field but with a significant temperature, density, and magnetic field difference between the two inflow regions. We began the analysis by noting that in the system under investigation, the flow stagnation point and electron current density peak do not coincide. As expected, the pressure nongyrotropy-based contribution to the reconnection electric field dominates at the flow stagnation point proper, but it also provides a large contribution at the current density peak.

We then proceeded to researching current and electron energy balance in the same way applied to a symmetric configuration earlier (Hesse et al., 2018). All components of the equations describing the time evolutions of electron current density and electron pressure were integrated around the current density peak and the flow stagnation point to determine the dominant contributions not only at the critical points, but beyond them. Similar to the symmetric system, we found that the nongyrotropic pressure terms generically conspire to reduce the current densities both at the stagnation point and the current density maximum. This result shows that also for asymmetric systems, the kinetic physics behind the pressure nongyrotropy is associated with current density dissipation. The reconnection electric field remains the key mechanism for sustaining the current density by accelerating particles to replenish those lost from the current density region. In addition to these two contributions we found, to a smaller degree around the stagnation point, and to a large degree around the current density maximum, an appreciable contribution of current density convection into or out of the integration volume, leading to a net negative contribution to the current balance at the current maximum. This, at first, surprising result could be explained by the prevailing electron in-plane flow patterns. Within those, the electron inflow against the *ExB* drift could be interpreted to result from a generalized diamagnetic drift, generated by the nongyrotropic pressure. In this sense, only one part of the pressure divergence serves to dissipate the current density, whereas another part is simply generating a diamagnetic drift-like plasma motion. These latter drifts do not generate significant contributions when integrated around the EDR of symmetric magnetic reconnection.

Regarding electron heating, we found here, like in symmetric reconnection, also that quasi-viscous contributions provide appreciable heating. The fact that the same nongyrotropic pressure terms, which contribute to current reduction, also play a role in electron heating, provides support for the intuition that dissipation of current-related, directed motion, ought to lead to an increase of random motion, i.e., thermal energy. Perhaps also as should be expected, the heating occurs around the gradients of the fastest electron flow velocity, which is associated with the main current density. Other components, e.g., related to shear-type gradients of the outflow velocity – similar to separatrix heating in symmetric reconnection, were found to be present in a neighborhood around the stagnation point as well, indicating that electron heating processes can operate at any suitable velocity shear layer.

Proceeding to an analysis of electron distributions, we found, in addition to previously described features (Chen et al., 2016b, Bessho et al., 2017, 2017), a number of new structures. In particular, we could reveal, around the stagnation point, a new, complex, crescent structure, which appears in crab leg-like shapes in reduced distributions, and which is clearly related to the orbital motion of electrons originating from the magnetospheric side of the EDR. In addition, we saw that these electrons, as well as those forming the main crescents, execute only a small number of bounces in the EDR, before they are expelled. In the case of the complex, accelerated distributions near the current density maximum, discrete substructures result from different particle origins and pre-acceleration in the magnetosheath inflow regions as well as in the Hall electric field component of the magnetospheric side. These particles are the main current carriers, and their rapid departure from the EDR leads to the macroscopic manifestation of current reduction. Similar to symmetric systems, their finite residence time suppresses plasma instabilities – in fact, standard linear instability theory does not apply here, which explains the laminar EDR structures observed by MMS (Burch et al., 2016).

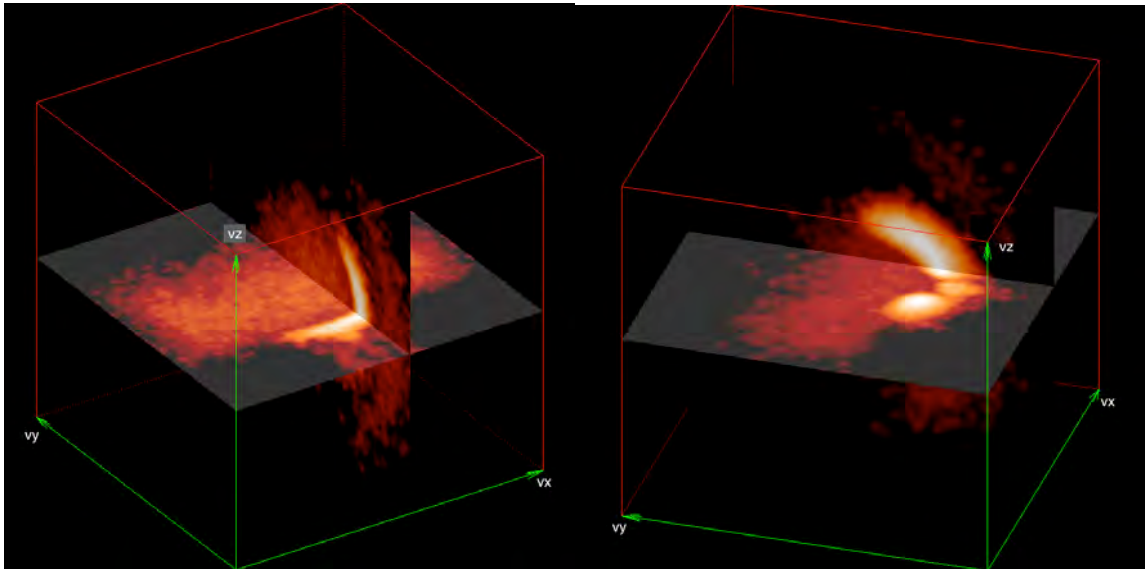

Figure 11 – Three-dimensional analysis of the 3D distributions behind F5c (left panel) and F5f (right panel), i.e., at the stagnation point and current density peak. We see that the crab legs visible in F5c are actually additional high-energy crescents, and that the complex, high-phase space density structures around the current density peak, are actually signatures of separate particle populations.

In summary, we found the conclusion that the reconnection electric field is a consequence of the need to maintain the current density in symmetric systems, also holds here. Complex particle orbital dynamics would lead, if no further acceleration were present, to a rapid outward diffusion of the electron current density. The electric field self-consistently adjusts itself to counter this loss effect. Future research should focus on extending these investigations to asymmetric systems with shear angles other than 180 degrees, and on, if possible, on finding analytic reconnection electric field models similar to the symmetric case (Hesse et al., 1999).

## 7. Acknowledgments

MH, TMJ, SFS, and HK acknowledge support by the University of Bergen, and MH and JB acknowledge support by NASA contract NNG04EB99C at SwRI. CN and PT were supported by NRC contract NN9496K. The simulation data are available at (Hesse, 2020).

## 8. References

Antiochos, S. K., C. R. DeVore, and J. A. Klimchuk, A model for solar coronal mass ejections, *Astrophys. J.,* 510, 485-493, 1999.

Bessho, N., L.-J. Chen, and M. Hesse (2016), Electron distribution functions in the diffusion region of asymmetric magnetic reconnection, *Geophys. Res. Lett.*, *43*, 1828–1836, doi:10.1002/2016GL067886.

Bessho, N, LJ Chen, M Hesse, S Wang (2017), The effect of reconnection electric field on crescent and U-shaped distribution functions in asymmetric reconnection with no guide field, Physics of Plasmas 24 (7), 072903

Burch, J. L., & Phan, T. D. (2016). Magnetic reconnection at the dayside magnetopause: Advances with MMS. *Geophys. Res. Lett.,* 43, 8327–8338. doi: 10.1002/2016GL069787

Burch, J. L., et al. (2016). Electron-scale measurements of magnetic reconnection in space. *Science,* 352(6290), aaf2939. doi: 10.1126/science.aaf2939

Cassak, P. A., & Shay, M. A. (2007). Scaling of asymmetric magnetic reconnection: General theory and collisional simulations. *Physics of Plasmas* 14, 102114 (2007). doi: 10.1063/1.2795630

Cassak, P. A., and Shay, M. A. (2009). Structure of the dissipation region in fluid simulations of asymmetric magnetic reconnection. *Physics of Plasmas,* 16, 055704. doi: 10.1063/1.3086867

Chen, L.-J., M. Hesse, S. Wang, N. Bessho, W. Daughton (2016a), Electron energization and structure of the diffusion region during asymmetric reconnection*, Geophys. Res. Lett.* ,43, 2405.

Chen, L.-J., et al. (2016b), Electron energization and mixing observed by MMS in the vicinity of an electron diffusion region during magnetopause reconnection, *Geophys. Res. Lett.* ,43, 6036.

Divin, A., G. Lapenta, S. Markidis, D. L. Newman, and M. V. Goldman (2012), Numerical simulations of separatrix instabilities in collisionless magnetic reconnection, Phys. Plasmas, 19, 042110.

Ergun, R. E., et al. (2016), Magnetospheric Multiscale Satellites Observations of Parallel Electric Fields Associated with Magnetic Reconnection, Phys. Rev. Lett. 116, 235102, *doi:10.1103/PhysRevLett.116.235102*.

Genestreti, K. J., et al. (2017). The effect of a guide field on local energy conversion during asymmetric magnetic reconnection: MMS observations. *J. Geophys. Res.: Space Physics*, 122, 11,342–11,353. doi: 10.1002/2017JA024247

Hesse, M., Aunai, N., Sibeck, D., & Birn, J. (2014). On the electron diffusion region in planar, asymmetric, systems. *Geophys. Res. Lett.*, 41, 8673–8680. doi: 10.1002/2014GL061586

Hesse, M., Schindler, K., Birn, J., & Kuznetsova, M. (1999). The diffusion region in collisionless magnetic reconnection. *Physics of Plasmas*, 6(5), 1781–1795. doi: 10.1063/1.873436

Hesse, M., et al. (2016). On the electron diffusion region in asymmetric reconnection with a guide magnetic field, *Geophys. Res. Lett.*, 43, doi: 10.1002/2016GL068373

Hesse, M., et al. (2018). The physical foundation of the reconnection electric field. *Physics of Plasmas,* 25, 032901. doi: 10.1063/1.5021461s

Hesse, M., & Cassak, P. A. (2020), Magnetic Reconnection in the Space Sciences: Past, Present, and Future. *Journal of Geophysical Research: Space Physics*, 124. https://doi.org/10.1029/2018JA025935

Hesse, Michael, 2020, "Replication Data for: A new Look at the Electron Diffusion Region in Asymmetric Magnetic Reconnection", https://doi.org/10.18710/Q4DIRK, DataverseNO

Liu, Y.-H., Hesse, M., Cassak, P. A., Shay, M. A., Wang, S., & Chen, L.-J. (2018). On the collisionless asymmetric magnetic reconnection rate. *Geophysical Research Letters*, 45, 3311– 3318. https://doi.org/10.1002/2017GL076460

Nakamura, R., et al. (2018), Structure of the current sheet in the 2017/07/11 electron diffusion region event, *J. Geophys. Res.*, 10.1029/2018JA026028

Nykyrii, K., and Otto, A. (2001), Plasma transport at the magnetospheric boundary due to reconnection in Kelvin-Helmholtz vortices, *Geophys. Res. Lett.*, 28, 3565–3568, doi: 10.1029/2001GL013239

Russell, C. T., and Elphic, R. C. (1979). ISEE observations of flux transfer events at the dayside magnetopause. *Geophys. Res. Lett.*, 6, 33-36. doi: 10.1029/GL006i001p00033

Shay, M. A., et al. (2016). Kinetic signatures of the region surrounding the X line in asymmetric (magnetopause) reconnection. *Geophys. Res. Lett.*, 43, 4145–4154. doi: 10.1002/2016GL069034

Uzdensky DA. (2011), Magnetic reconnection in extreme astrophysical environments. *Space Sci. Rev.* 160, 45–71, *doi:10.1007/s11214-011-9744-5*